\documentclass[sigconf,natbib=true,anonymous=false]{acmart}

\makeatletter
\def\@ACM@checkaffil{
    \if@ACM@instpresent\else
    \ClassWarningNoLine{\@classname}{No institution present for an affiliation}%
    \fi
    \if@ACM@citypresent\else
    \ClassWarningNoLine{\@classname}{No city present for an affiliation}%
    \fi
    \if@ACM@countrypresent\else
        \ClassWarningNoLine{\@classname}{No country present for an affiliation}%
    \fi
}
\makeatother

\setcopyright{none}
\renewcommand\footnotetextcopyrightpermission[1]{} 
\def\CHK#1 {\textcolor{magenta}{{\bf [CHK:}~#1{\bf ]}}~}
\def\ADD#1 {\textcolor{cyan}{{\bf [ADD:}~#1{\bf}]}~}

\usepackage[utf8]{inputenc}

\usepackage{subcaption} 

\usepackage{multirow} 

\usepackage{amsmath,empheq} 

\usepackage[ruled,linesnumbered]{algorithm2e} 
\SetKwComment{Comment}{/* }{ */} 

\usepackage{enumitem}
\setlist[itemize]{leftmargin=*}
\setlist[enumerate]{leftmargin=*}

\renewcommand{\figureautorefname}{Figure}

\def\equationautorefname~#1\null{Eqn. ~(#1)\null}
\def\figureautorefname~#1\null{Fig. ~#1\null}

\newtheoremstyle{remboldstyle}
  {}{}{\itshape}{}{\bfseries}{:}{.5em}{{\thmname{#1 }}{\thmnumber{#2.}}{\thmnote{ #3}}}
\theoremstyle{remboldstyle}

\acmConference[MM'22]{ACM MM'22: ACM International Conference on Multimedia}{October 10-14, 2022}{Lisbon, Portugal}
\acmBooktitle{MM'22: ACM International Conference on Multimedia, October 10-14, 2022}
\acmBooktitle{ }

\copyrightyear{2022} 
\acmYear{2022} 
\setcopyright{acmcopyright}\acmConference[MM '22]{Proceedings of the 30th ACM International Conference on Multimedia}{October 10--14, 2022}{Lisboa, Portugal}
\acmBooktitle{Proceedings of the 30th ACM International Conference on Multimedia (MM '22), October 10--14, 2022, Lisboa, Portugal}
\acmPrice{15.00}
\acmDOI{10.1145/3503161.3548098}
\acmISBN{978-1-4503-9203-7/22/10}

\begin{document}

\begingroup
\hyphenpenalty 9000
\exhyphenpenalty 9000

\title{Micro-video Tagging via Jointly Modeling Social Influence and Tag Relation}

\author{Xiao Wang}
\email{scz.wangxiao@gmail.com}
\affiliation{%
  \institution{Shandong University}
}

\author{Tian Gan}
\email{gantian@sdu.edu.cn}
\authornote{Corresponding authors: Tian Gan and Jianlong Wu.}
\affiliation{%
  \institution{Shandong University}
}

\author{Yinwei Wei}
\email{weiyinwei@hotmail.com}
\affiliation{%
  \institution{National University of Singapore}
}

\author{Jianlong Wu}
\email{jlwuhust@gmail.com}
\authornotemark[1]
\affiliation{%
  \institution{Harbin Institute of Technology (Shenzhen) \& Shandong University}
}

\author{Dai Meng}
\email{daimeng@kuaishou.com}
\affiliation{%
  \institution{Kuaishou Technology}
}

\author{Liqiang Nie}
\email{nieliqiang@gmail.com}
\affiliation{%
  \institution{Harbin Institute of Technology (Shenzhen)}
}

\begin{abstract}
The last decade has witnessed the proliferation of micro-videos on various user-generated content platforms. 
According to our statistics, around 85.7\% of micro-videos lack annotation. 
In this paper, we focus on annotating micro-videos with tags.
Existing methods mostly focus on analyzing video content, neglecting users' social influence and tag relation.
Meanwhile, existing tag relation construction methods
suffer from either deficient performance or low tag coverage.
To jointly model social influence and tag relation, 
we formulate micro-video tagging as a link prediction problem in a constructed heterogeneous network.
Specifically, the tag relation (represented by tag ontology) is constructed in a semi-supervised manner.
Then, we combine tag relation, video-tag annotation, and user follow relation to build the network.
Afterward, a better video and tag representation are derived through Behavior Spread modeling and visual and linguistic knowledge aggregation. 
Finally, the semantic similarity between each micro-video and all candidate tags is calculated in this video-tag network.
Extensive experiments on industrial datasets of three verticals verify the superiority of our model compared with several state-of-the-art baselines. 
\end{abstract}

\ccsdesc[500]{Information systems~Document representation}


\keywords{Micro-video Tagging; Behavior Spread; Ontology Construction}

\maketitle

\vspace{-1em}
\section{Introduction}
The last decade has evidenced the prosperity of micro-videos in User-Generated Content (UGC) platforms. 
To strengthen the applications like searching and recommendation \cite{wei2019neural, wei2020graph, sun2022response}, tags are widely-used to summarize micro-videos.
Considering the fact that users may not add sufficient tags when uploading micro-videos, tagging has hence become an expensive routine for operation teams on UGC platforms. 
According to the statistics over 600 million videos collected from Kuaishou platform, around 85.7\% of them have no tags at all. 
In order to generate tags with minimal human efforts, automatic micro-video tagging has drawn considerable attention from industrial and academic communities. 
Most existing methods formulate video tagging as a multi-label classification problem. 
Basic methods leverage information from video content and descriptions for tagging \cite{nextvlad, wu_tencent_mmads_1st_2021}. 
Sophisticated methods leverage extra information, such as tag graph \cite{ml_gcn, cma, li_mall_2022, jin_transfusion_2021}, query log \cite{taggnn}, user behavior \cite{jit2r}, and user profile \cite{wei_personalized_2019, li_long-tail_2019} to assist tagging. 

\begin{figure}[t]
    \centering
    \includegraphics[width=0.95\linewidth]{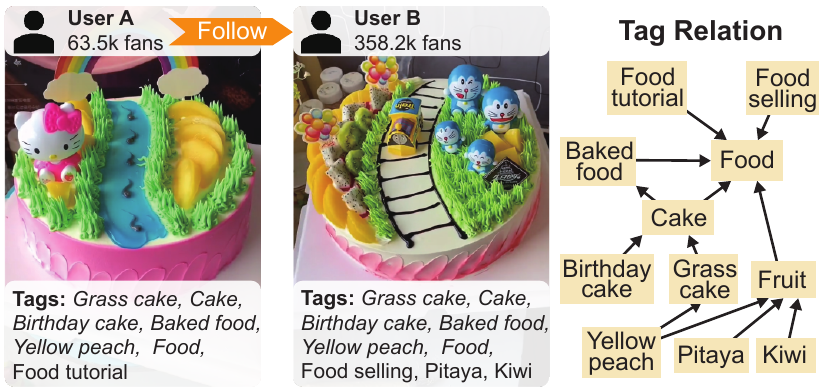}
    \vspace{-1em}
    \caption{
    Example of social influence and tag relation. Social influence: user A is a follower of B, and imitates user B to create a video with similar tags (illustrated in \textit{italics}).}
    \vspace{-2em}
    \label{fig:task_example}
\end{figure}

However, users’ social influence and tag relation are rarely discussed, which could have played a pivotal role. 
On one hand, users’ social influence is crucial since user may imitate their social network neighbors to create similar micro-videos, especially on UGC platforms because users are not only content creators but also consumers. 
%
Taking \autoref{fig:task_example} as an example, user B produces a micro-video about a ``grass cake'', 
and then the follower, user A, imitates this video to produce a similar one with almost the same set of tags. 
Such imitation phenomenon in the social network, namely \textbf{Behavior Spread} \cite{centola2010spread}, can enhance tagging performance if correctly modeled.
On the other hand, tag relation is reflected by a \textbf{tag ontology},
which is often represented as a Directed Acyclic Graph (DAG) composed of tags with \textit{is\_subtopic\_of} relations. 
Tag ontology facilitates tagging by providing external knowledge from three aspects: (i) tag dependencies \cite{cma, kssnet}, (ii) top-down knowledge transfer \cite{li_long-tail_2019}, and (iii) buttom-up semantics abstraction \cite{nie_large-scale_2020}.

To incorporate users’ social influence and tag relations, we build a heterogeneous user-video-tag network.
This heterogeneous graph neural network is then used to derive a better video and tag representation.
Consequently, we are able to measure the video-tag similarly to get tagging results.
%
%
There are still challenges for the above approach:
\textbf{C1: Behavior Spread modeling}. 
We need to propose an effective model to filter irrelevant information for Behavior Spread modeling, since not all micro-videos are created through imitation.
\textbf{C2: Visual-linguistic knowledge aggregation}. 
There are two sources of knowledge for learning a tag representation: visual knowledge from videos and linguistic knowledge from relevant tags in the tag ontology. 
Existing aggregation methods \cite{kssnet, cma} that apply direct aggregation approaches like vector concatenation or attention mechanism are sub-optimal due to the ignorance of the redundancy in common knowledge. 
Concretely, concatenation duplicates the common knowledge, and attention mechanism tends to over concern the unique knowledge \cite{lu_aan_ref_ReID_2020, li_aan_ref_CTR_2020, verma_deepcu_2019}. 
\textbf{C3: Tag ontology construction}. 
Most of the methods \cite{kssnet, li_long-tail_2019} acquire the desired tag ontology from existing open knowledge bases. However, because video content in UGC platforms changes very fast, open knowledge bases can cover only a portion of tags. Other methods \cite{ml_gcn, fang_folksonomy-based_2016} attempt to build the tag ontology from tag statistics. These methods are rule-based and thus lack accuracy.


To solve \textbf{C3}, we construct a tag ontology using semi-supervised classification based upon hand-crafted features. 
%
%
After that, we design an adve\textbf{R}sarial aggregated g\textbf{A}te\textbf{D} gr\textbf{A}ph t\textbf{R}ansformer network (\textbf{RADAR}) to propagate information over the entire graph 
for a better representation of each micro-video and tag node.
RADAR is a heterogeneous Graph Neural Network (GNN) composed of a Gated Graph Transformer (GGT) and an Adversarial Aggregation Network (AAN). 
GGT aims to tackle \textbf{C1} by aggregating information from neighbors, and filtering irrelevant information. 
To cope with \textbf{C2}, AAN is applied to aggregate visual and linguistic knowledge by removing the redundant information while keeping the complementary. 
Finally, we compute the semantic similarity between a given micro-video and all candidate tags as our predictions.
To justify our model, we conduct extensive experiments over large-scale real-world datasets. The experimental results demonstrate the effectiveness of GGT and ANN, and our methods are consistently superior to several start-of-the-art baselines.

In summary, the contributions of this work are threefold:
\begin{itemize}
    \item We construct a dataset of 450,000 micro-videos with their creators' social network in a UGC platform, and our experiments show that social network can benefit video tagging. To the best of our knowledge, this is the first work on modeling users' social influence towards micro-video tagging.
    \item We propose a semi-supervised manner to build a tag ontology, outperforming existing methods with little human effort.
    \item We design RADAR, a heterogeneous GNN composed of GGT and AAN to jointly model users’ social influence and tag relation in an end-to-end manner. RADAR outperforms cutting-edge methods with a clear margin in real-world datasets. Our code is available\footnote{\url{https://github.com/SCZwangxiao/RADAR-MM2022.git}}.
\end{itemize}
\vspace{-1em}
\section{Related Work}

\subsection{Video Tagging}
Video tagging aims to find keywords that can describe the core content of a video. All methods can be categorized into two types based on the information they use.
\textbf{Basic methods} leverage information only from video content: textual, visual, and audio. There are two sub-types. Key-phrase extraction methods extract tags from video descriptions \cite{joint_kpe, autophrase, tag_from_danmu}, and rank them according to their relevance with core semantics \cite{joint_kpe, kotkov_tagranking_2021}. Multi-label classification methods assign tags from a predefined word set to videos \cite{nextvlad, hvu, wu_tencent_mmads_1st_2021}. 
\textbf{Sophisticated methods} leverage extra information besides video content, such as tag graph with knowledge \cite{ml_gcn, cma, li_mall_2022, jin_transfusion_2021}, query log \cite{taggnn}, user behavior \cite{jit2r}, and user profile \cite{wei_personalized_2019, li_long-tail_2019}.
Among all the above extra information, tag graph
has attracted the most attention. Generally, the tag graph is built either from tag co-occurrence \cite{ml_gcn, cma, kssnet} or common knowledge bases \cite{li_zeroshot_2015, kssnet, li_long-tail_2019}. However, the former can only provide statistical relations, and the latter has low tag coverage. Contrary to the tag graph, social network information (user follow, specifically) has not been touched in video tagging.

\vspace{-1em}
\subsection{Heterogeneous Graph Neural Network}
Heterogeneous networks are graphs with more than one meta-paths. 
Heterogeneous graph neural networks are widely applied to derive a better node representation for them \cite{zhang_heteGNNref1_2021, liu_heteGNNref2_2021, zhang_heteGNNref3_2021, fan_heteGNNref4_2021}. For example, \citeauthor{han} \cite{han} proposed a method which is based on two-level attention. Concretely, node-level attention measures the importance of neighbors within each meta-path, and semantic-level attention measures the importance of different meta-paths. \citeauthor{hgt} \cite{hgt} proposed a method in which node and edge-type dependent parameters are used to parameterize the graph transformer network of each meta-path. \citeauthor{simple_hgn} \cite{simple_hgn} proposed a method to demonstrate that a simple graph attention network combined with three well-known techniques can outperform all previous models.

\vspace{-1em}

\begin{figure*}[t]
    \centering
    \includegraphics[width=0.99\linewidth]{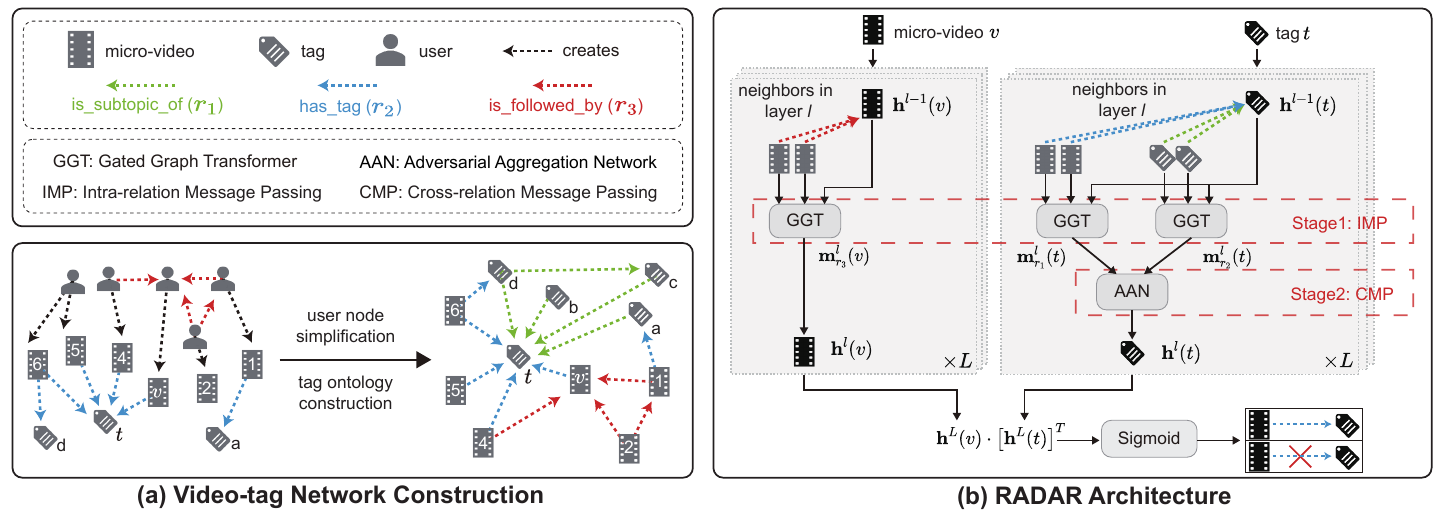}
    \vspace{-1em}
    \caption{Method overview. (a) The video-tag network is built from the raw data composed of users and their micro-videos annotated with tags. (b) The GNN propagation of our RADAR network in each layer has two stages.}
    \vspace{-1em}
    \label{fig:overview}
\end{figure*}

\section{Problem Formulation}
We formulate micro-video tagging as a link prediction problem on a video-tag network $\boldsymbol{G}(\mathcal{V}, \mathcal{E})$. $\mathcal{V}$ is the node set composed of videos and tags, and $\mathcal{E}$ is the edge set. Note that video tagging with graphical data can also be formulated as a knowledge graph embedding problem \cite{jin_transfusion_2021}. However, such formulation only supports transductive learning, and is unable to handle new videos. 

For each node $u\in\mathcal{V}$, we denote its initial representation as $\mathbf{h}^0(u)\in\mathbb{R}^d$, where $d$ is the size of hidden dimension. For tag node, $\mathbf{h}^0$ comes from the sum of word embedding and learnable embedding. For video node, $\mathbf{h}^0$ comes from aggregated frame embedding. Given $\mathbf{h}^0$ and $\boldsymbol{G}$, our RADAR model, a $L$-layer heterogeneous GNN, can derive a better node representation $\mathbf{h}^L\in\mathbb{R}^d$:
\begin{equation} \label{eq:RADAR_formulation}
    \mathbf{h}^L = \textmd{RADAR} \left( \mathbf{h}^0, \boldsymbol{G} \right).
\end{equation}
Based on this representation, we predict the confidence score $\hat{y}(v,t)\in\mathbb{R}$ that micro-video $v\in\mathcal{V}_{\textmd{video}}$ has tag $t\in\mathcal{V}_{\textmd{tag}}$:
\begin{equation}
    \hat{y}(v,t) = \textmd{Sigmoid} \left( \mathbf{h}^L(v) \cdot \left [ \mathbf{h}^L(t) \right ]^T  \right).
\end{equation}


\section{Methodology} \label{sec:method}

Our overall framework is summarized in \autoref{fig:overview}. We first propose a semi-supervised classification method to build a tag ontology, solving the ontology construction challenge. We then build the heterogeneous video-tag network for jointly modeling social influence and tag ontology. Afterward, we design the adve\textbf{R}sarial aggregated g\textbf{A}te\textbf{D} gr\textbf{A}ph t\textbf{R}ansformer network (RADAR) to derive a better video and tag representation. The two components of RADAR solve Behavior Spread modeling and visual-linguistic knowledge aggregation challenges respectively. Finally, we estimate the semantic similarity between a micro-video and all tags as our predictions.

\vspace{-1em}

\subsection{Tag Ontology Construction} \label{sec:ontology_construction}
In this subsection, we construct the tag ontology, a graph composed of tags and \textit{is\_subtopic\_of} relations. It is naturally a Directed Acyclic Graph (DAG), because a cycle made of \textit{is\_subtopic\_of} relations would cause a paradox. Ontology construction consists of two steps: subtopic discovery and DAG construction, as summarized in \autoref{fig:ontology_construction}. In subtopic discovery, we discover subtopic relations among tags using semi-supervised classification based on hand-crafted features. These features are derived from tag statistics in video-tag annotations so that we can cover all tags. This step will produce a DAG violated graph. Following that, graphical constraints are imposed to derive a DAG as the final tag ontology. 



\begin{figure*}[t]
    \includegraphics[width=0.9\linewidth]{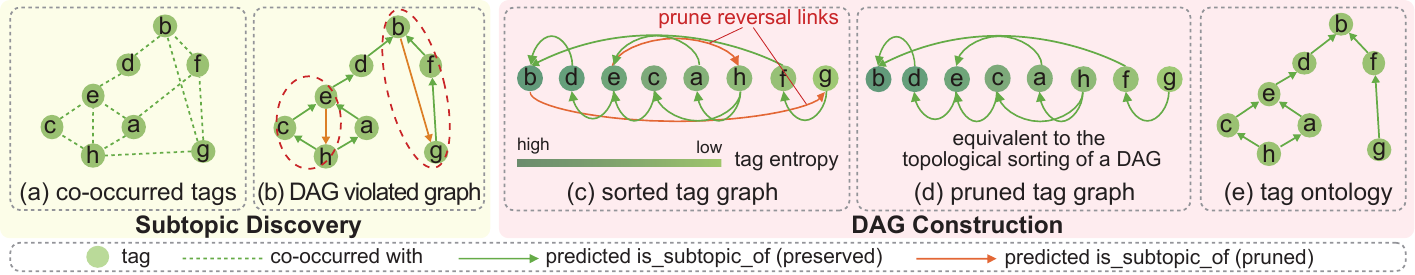}
    \vspace{-1em}
    \caption{Tag Ontology Construction. (a)(b) \textit{Is\_subtopic\_of} relations are discovered, and form a DAG violated graph. (c) All tags are sorted, and then the reversal links are pruned. (d)(e) The pruned tag graph is equivalent to a DAG, which is our tag ontology.}
    \label{fig:ontology_construction}
\end{figure*}

\subsubsection{Subtopic Discovery} \label{subtopic_discovery}

We presume that \textit{if two tags have subtopic relations, they will co-occur in at least one micro-video}. Based on this, we formulate subtopic discovery as a binary classification problem on co-occurred tags. Formally, for each co-occurred tag pair $(u,v)$, $u,v\in\mathcal{V}_{\textmd{tag}}$, we predict their subtopic score $r(u,v)$:
\begin{align}
    r(u,v) &= g\left(\mathbf{k}(u,v)\right), \label{eq:sub_dis_classification} \\
    r(u,v) &=
        \begin{cases}
            1 & \text{ if } \textmd{$v$ is a subtopic of $u$}, \\
            0 & \text{ if } \textmd{no relationship between $u$ and $v$},
        \end{cases}
\end{align}
where $g(\cdot)$ can be an arbitrary classifier, $\mathbf{k}(u,v)\in\mathbb{R}^{d_r}$ is our hand-crafted features, and $d_r$ is the dimension of feature. The features are designed to reflect the following two properties between tags:

\textbf{Semantic overlap} refers to the overlap of meaning between two tags. If two tags $u$ and $v$ have subtopic relations, they must have semantic overlap. We measure the semantic overlap using Point-wise Mutual Information $\textmd{PMI}(u,v)\in\mathbb{R}$:
\begin{equation}
    \textmd{PMI}(u,v) = \textmd{log}\left ( \frac{p(u,v)}{p(u)p(v)} \right ),
\end{equation}
and Point-wise Kullback-Leibler divergence $\textmd{PKL}(u,v)\in\mathbb{R}$:
\begin{equation}
    \textmd{PKL}(u,v) = p(u,v)\textmd{log}\left ( \frac{p(u,v)}{p(u)p(v)} \right ),
\end{equation}
where $p(u,v)$ is the probability that $u,v$ occur in the same micro-video, and $p(u), p(v)$ are the occurrence probability of $u,v$, respectively. Compared with PMI, PKL is less biased towards the rare-occurred tags due to the additional $p(u,v)$.

\textbf{Semantic broadness} means the broadness of tag meaning. Broader semantics indicates a higher level in tag ontology (e.g., ``food'' has broader semantics than ``cake''). We measure it using two features. One is tag transfer probability $p(u|v)\in\mathbb{R}$ from $v$ to $u$:
\begin{equation}
    p(u|v) = \frac{\left |\textmd{videos that have tag $u$ and $v$}\right | }{\left |\textmd{videos that have tag $v$}\right | }.
\end{equation}
Intuitively, the larger $p(u|v)$ is, the more likely that $v$ is a subtopic of $u$. For example, given that tag ``cake'' is a subtopic of ``food'', and ``cake'' is used, it will be very likely to see ``food'' as well (i.e., large $p(\textmd{food}|\textmd{cake})$). Empirically, this probability has proven to be effective in industrial concept mining system \cite{liu_concept_2019}. The other feature is tag entropy $H(u)\in\mathbb{R}$:
\begin{equation}
    H(u) = - \sum_{w}{p(u|w)\textmd{log}(p(u|w))}.
\end{equation}
Intuitively, a tag with higher $H(u)$ has more and stronger inbound tag transfers from others, indicating a higher level in tag ontology. $H(u)$ is also used by FBVO \cite{fang_folksonomy-based_2016} to derive visual ontology.

With the semantic overlap and broadness features above, we can define our hand-crafted feature $\mathbf{k}(u,v)$ containing six basic features: $p(u|v)$, $p(v|u)$, $H(u)$, $H(v)$, $PMI(u,v)$, $PKL(u,v)$, and two second order features: $log\left ( p(u|v)/p(v|u) \right )$, $log\left ( H(u)/H(v) \right )$. Based upon these, the classification in \autoref{eq:sub_dis_classification} is in a semi-supervised manner. Concretely, the labels come from a small portion of the co-occurred tag pairs. Meanwhile, the features can be computed from all co-occurred tags to leverage information in unlabeled data.

For each tag pair ($u,v$), we can estimate the confidence score of $v$ being a subtopic of $u$: $\hat{r}(u,v)$. We keep those tag pairs whose $\hat{r}$ are larger than a threshold $\delta_r$, and link them into a directed graph.

\subsubsection{DAG Construction}
Since not all predictions are correct, there will be DAG violations definitely, as exemplified in \autoref{fig:ontology_construction}(b). Thus, we impose graphical constraints to derive the final DAG. 

As illustrated in \autoref{fig:ontology_construction}(c), we first sort all tags according to their entropy $H$ in descending order. Considering higher entropy means a higher level in tag ontology, we then keep \textit{is\_subtopic\_of} relations only from tags with lower entropy to higher ones. Afterward, the sorted tag list can be seen as the topological sorting of a DAG, which can be our tag ontology. Note that there might be isolated components, because lots of edges are filtered when we keep $\hat{r}(u,v)>\delta_r$ in subtopic discovery. For those components, we relax the constraint until they are not isolated or $\hat{r}(u,v)< \epsilon_r$, and link the remaining isolated ones to the tag with the highest entropy.

\subsection{Video-tag Network} \label{sec:video_tag_network}
We build a directed heterogeneous video-tag network to incorporate social network and tag ontology, as shown in \autoref{fig:overview}(a).

Our raw data is composed of users, micro-videos, and tags along with two relations: \textit{has\_tag} and \textit{is\_followed\_by}. The latter provides a clue about Behavior Spread. 

We first simplify the raw data by deleting user nodes, and letting their created videos inherit \textit{is\_followed\_by} relations. Concretely, if $\textmd{user}_A$ is followed by $\textmd{user}_B$, then all $\textmd{user}_A$'s videos are followed by $\textmd{user}_B$'s videos. Note that we only preserve the \textit{is\_followed\_by} relations pointing from previously-uploaded videos to newly-uploaded ones, because only old videos can influence new ones. After simplification, we incorporate the \textit{is\_subtopic\_of} relations from the constructed tag ontology to get the video-tag network.

Formally, we denote the micro-video set as $\mathcal{V}_{\textmd{video}}$ and tag set as $\mathcal{V}_{\textmd{tag}}$. They form the node set $\mathcal{V} = \mathcal{V}_{\textmd{video}} \cup \mathcal{V}_{\textmd{tag}}$. For convenience, we denote the three relations \textit{is\_subtopic\_of}, \textit{has\_tag}, and \textit{is\_followed\_by} as $r_1,r_2$, and $r_3$, respectively. Their corresponding relation sets are $\mathcal{E}_{r_1}$, $\mathcal{E}_{r_2}$, $\mathcal{E}_{r_3}$, respectively. They form the edge set $\mathcal{E} = \mathcal{E}_{r_1} \cup \mathcal{E}_{r_2} \cup \mathcal{E}_{r_3}$.

\subsection{RADAR Architecture} \label{sec:method}

\subsubsection{Overall RADAR Architecture} \label{sec:RADAR_overview}
To derive a better video and tag representation, we design RADAR, an $L$ layer graph neural network. Considering that tag nodes have two types of inbound edges, while video nodes have only one, we apply different GNN propagation strategies to them. Concretely, we divide propagation into two stages: \textit{intra-relation message passing} for both video and tag nodes and \textit{cross-relation message aggregation} for tag nodes only. Correspondingly, each GNN layer in RADAR has two components: \textbf{Gated Graph Transformer} (GGT) and \textbf{Adversarial Aggregation Network} (AAN), as illustrated in \autoref{fig:overview}(b).

\begin{figure*}[t]
    \centering
    \includegraphics[width=0.8\linewidth]{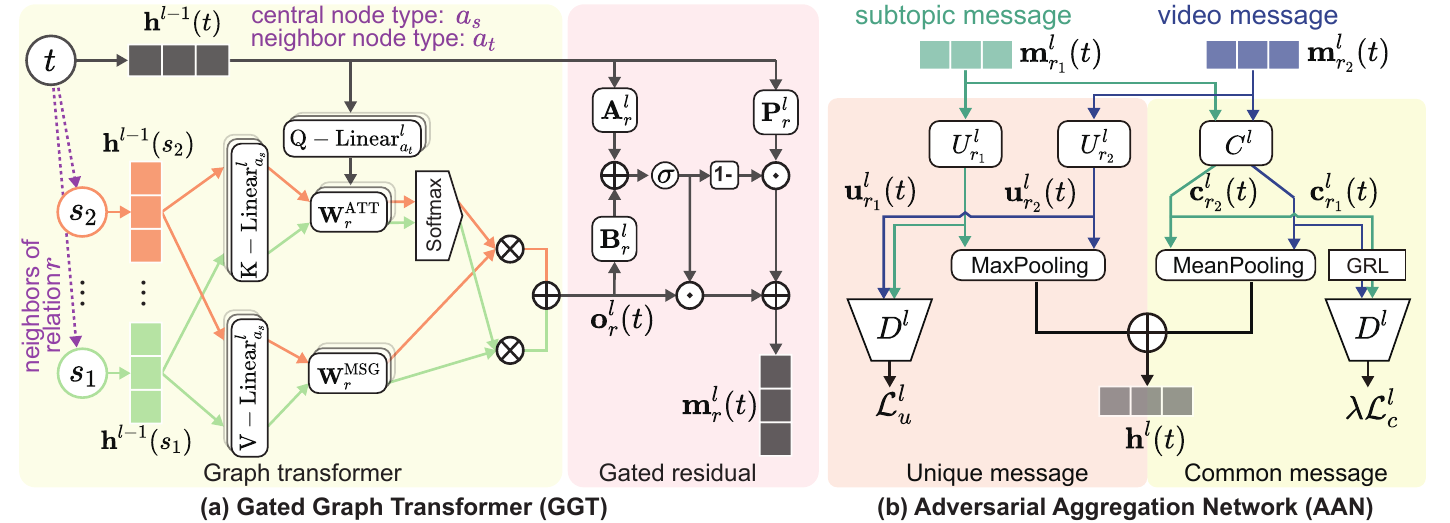}
    \vspace{-1em}
    \caption{(a) GGT aggregates neighbor information using graph transformer, and filters irrelevant information using gated residual network.
    (b) ANN separates subtopic and video messages into common and unique information, and aggregates them.}
    \label{fig:ggt_aan}
\end{figure*}

In the \textit{intra-relation message passing} stage, GGT is applied for both tag and video nodes to derive message information propagated from their neighbors. Formally, video $v$ receives message $\mathbf{m}_{r_3}^l(v)\in\mathbb{R}^d$ passed from its $r_3$ neighbors:
\begin{equation} \label{eq:follow_msg}
    \mathbf{m}_{r_3}^l(v) = \textmd{GGT}\left( \mathbf{h}^{l-1}(v), \left\{\mathbf{h}^{l-1}(s)|s\in\mathcal{N}^l_{r_3}(v)\right\} \right),
\end{equation}
and video representation is updated by:
\begin{equation} \label{eq:follow_msg}
    \mathbf{h}^l(v) = \mathbf{m}_{r_3}^l(v),
\end{equation}
where the superscript $l$ denotes the ($l$)-th layer, $\mathbf{h}^l(v)\in\mathbb{R}^d$ is the output representation of node $v$, and $\mathcal{N}^l_{r_3}(v)$ are node $v$'s neighbors of $r_3$ relations. 
Meanwhile, tag $t$ receives messages $\mathbf{m}_{r_1}^l(t)$, $\mathbf{m}_{r_2}^l(t)\in\mathbb{R}^d$ passed from its $r_1$ and $r_2$ neighbors, respectively:
\begin{empheq}[left=\empheqlbrace]{align} \label{eq:two_message_tag_node}
    &\mathbf{m}_{r_1}^l(t) = \textmd{GGT}\left( \mathbf{h}^{l-1}(t), \left\{\mathbf{h}^{l-1}(s)|s\in\mathcal{N}^l_{r_1}(t)\right\} \right),\\
    &\mathbf{m}_{r_2}^l(t) = \textmd{GGT}\left( \mathbf{h}^{l-1}(t), \left\{\mathbf{h}^{l-1}(s)|s\in\mathcal{N}^l_{r_2}(t)\right\} \right),
\end{empheq}
where $\mathcal{N}^l_{r_1}(t), \mathcal{N}^l_{r_2}(t)$ are node $t$'s neighbors of $r_1$ and $r_2$ relations, respectively.

For each video, the neighborhood messages carry information about the social influence of its \textit{is\_followed\_by} neighbors. Since not all videos are created through imitation, in GGT we use a gated mechanism to filter irrelevant information for tackling Behavior Spread modeling challenge. 
For each tag $t\in\mathcal{V}_{\textmd{tag}}$, the neighborhood messages bring linguistic and visual knowledge from \textit{is\_subtopic\_of} and \textit{has\_tag} neighbors, respectively. Therefore, we propose AAN to solve the visual-linguistic aggregation challenge in the next stage.

In the \textit{cross-relation message aggregation} stage, AAN is applied for tag nodes only to aggregate subtopic messages $\mathbf{m}_{r_1}^l(t)$ and video messages $\mathbf{m}_{r_2}^l(t)$. For each tag $t\in\mathcal{V}_{\textmd{tag}}$, its representation is updated by:
\begin{equation} \label{eq:aggregation}
    \mathbf{h}^l(t) = \textmd{AAN}\left(\mathbf{m}^l_{r_1}(t), \mathbf{m}^l_{r_2}(t)\right).
\end{equation}

After updating node representations following the above two stages in each GNN layer, we get the final representation $\mathbf{h}^L$.

\subsubsection{Gated Graph Transformer} \label{sec:ggt}

The goal of GGT is to derive a message information $\mathbf{m}^l_r(t)\in\mathbb{R}^d$ propagated from their neighbors in the ($l$)-th layer, where $r$ is the neighbor relation. As illustrated in \autoref{fig:ggt_aan}(a), we apply graph transformer to aggregate neighborhood information, and gated residual network to filter irrelevant information. Formally, we denote the central node as target $t$ and neighbor node as source $s\in\mathcal{N}^l_r(t)$. The GGT is parameterized by target node type $a_t$, source node type $a_s$, and neighbor relation $r$.

Inspired by the Transformer architecture \cite{transformer, hgt}, we first treat target node $t$ as the query $\mathbf{q}_r^l(t)\in\mathbb{R}^d$, and source nodes $s\in\mathcal{N}^l_r(t)$ as both keys $\mathbf{k}_r^l(s)\in\mathbb{R}^d$ and values $\mathbf{v}_r^l(s)\in\mathbb{R}^d$:
\begin{empheq}[left=\empheqlbrace]{align}
    &\mathbf{q}_r^l(t) = \textmd{Q-Linear}^l_{a_t} \left( \mathbf{h}^{l-1}(t) \right),\\
    &\mathbf{k}_r^l(s) = \textmd{K-Linear}^l_{a_s} \left( \mathbf{h}^{l-1}(s) \right),\\
    &\mathbf{v}_r^l(s) = \textmd{V-Linear}^l_{a_s} \left( \mathbf{h}^{l-1}(s) \right)\mathbf{W}_r^{\textmd{MSG}},
\end{empheq}
where $\textmd{Q-Linear}^l_{a_t}$, $\textmd{K-Linear}^l_{a_s}$, $\textmd{V-Linear}^l_{a_s}$ are the linear transformation layers corresponding to specific node types, and $\mathbf{W}_r^{\textmd{MSG}}\in\mathbb{R}^{d\times d}$ is the message matrix of relation $r$.

Then, we calculate dot product between query and keys, and apply softmax to measure the importance of source node $s$ with respect to target node $t$ as $\alpha_r^l(t,s)\in\mathbb{R}$:
\begin{equation} \label{graph_attention}
    \alpha_r^l(t,s) = \frac{exp \left( \mathbf{q}_r^l(t)\mathbf{W}_r^{\textmd{ATT}}\mathbf{k}_r^l(s)^T / \sqrt{d} \right)}
    {\sum_{w\in\mathcal{N}^l_r(t)}{exp \left(\mathbf{q}_r^l(t)\mathbf{W}_r^{\textmd{ATT}}\mathbf{k}_r^l(w)^T / \sqrt{d} \right)}},
\end{equation}
where $\mathbf{W}_r^{\textmd{ATT}}\in\mathbb{R}^{d\times d}$ is an edge-based projection matrix. Note that the softmax is done within source nodes of the same relation $r$, and we name it separate attention. This is different from the mutual attention in the existing graph transformer \cite{hgt} in which normalization is done within neighbors of all relations. We adopt separate attention because the different relations in the video-tag network are substantially different, and can not be weighed together.

Afterward, we can get the output of graph transformer $\mathbf{o}^l(t)\in\mathbb{R}^d$ by weighted summation of values:
\begin{equation}
    \mathbf{o}_r^l(t) = \sum_{s\in\mathcal{N}^l_r(t)}{\alpha_r^l(t,s)\mathbf{v}_r^l(s)}.
\end{equation}

Finally, we propose the \textbf{gated residual network} to derive neighborhood message $\mathbf{m}^l_r(t)$:
\begin{empheq}[left=\empheqlbrace]{align}
    &\mathbf{z}_r^l(t) = \textmd{Sigmoid} \left(
    \mathbf{A}^l_r \mathbf{h}^{l-1}(t) + 
    \mathbf{B}^l_r \mathbf{o}_r^l(t)
    \right), \\
    &\mathbf{m}_r^l(t) =  \mathbf{z}_r^l(t) \odot \mathbf{o}_r^l(t) + 
    \left(\mathbf{1}-\mathbf{z}_r^l(t)\right) \odot \left(\mathbf{P}^l_r \mathbf{h}^{l-1}(t) \right ),
\end{empheq}
where $\mathbf{z}_r^l(t)\in\mathbb{R}^d$ is the gating values, $\odot$ denotes element-wise multiplication, and $\mathbf{A}^l_r$, $\mathbf{B}^l_r$, $\mathbf{P}^l_r\in\mathbb{R}^{d\times d}$ are linear transform matrices with respect to relation $r$ in the ($l$)-th layer. The gated mechanism generally filters irrelevant information. For \textit{is\_followed\_by} neighbors, it suppresses neighbors when videos are created not from imitation. For \textit{is\_subtopic\_of} neighbors, it keeps only general semantics as an abstraction of subtopic tags.

For simplicity, we omit GELU activation function after $\mathbf{P}^l_r$, bias vector after $\mathbf{A}^l_r$, $\mathbf{B}^l_r$, and $\mathbf{P}^l_r$, and superscript $l$ in $\mathbf{W}_r^{\textmd{MSG}}$ and $\mathbf{W}_r^{\textmd{ATT}}$.

\subsubsection{Adversarial Aggregation Network} \label{sec:aan}

The AAN is designed only for tag nodes to aggregate linguistic and visual knowledge from subtopic message and video messages, respectively. Both messages can be divided into unique (exists in one), and common information (exists in both). Existing methods such as concatenation or attention mechanism \cite{han} are sub-optimal due to the ignorance of the redundancy in common information. Concretely, concatenation duplicates the common information, and attention mechanism tends to overlook the unique information \cite{lu_aan_ref_ReID_2020, li_aan_ref_CTR_2020}. 

In our AAN, we propose to separate common and unique information in order to remove the redundant information while keeping the complementary one. Formally, for each target tag node $t\in\mathcal{V}_{\textmd{tag}}$, the incoming messages $\mathbf{m}_{r_1}^l(t)$ and $\mathbf{m}_{r_2}^l(t)$ are first separated into common message $\mathbf{c}_{r_1}^l(t)\in\mathbb{R}^d$, $\mathbf{c}_{r_2}^l(t)\in\mathbb{R}^d$:
\begin{empheq}[left=\empheqlbrace]{align}
    &\mathbf{c}_{r_1}^l(t) = C^l \left(
        \mathbf{m}_{r_1}^l(t)
    \right), \\
    &\mathbf{c}_{r_2}^l(t) = C^l \left(
        \mathbf{m}_{r_2}^l(t)
    \right),
\end{empheq}
and unique message $\mathbf{u}_{r_1}^l(t)$, $\mathbf{u}_{r_2}^l(t)$:
\begin{empheq}[left=\empheqlbrace]{align}
    &\mathbf{u}_{r_1}^l(t) = U^l_{r_1} \left(
        \mathbf{m}_{r_1}^l(t)
    \right), \\
    &\mathbf{u}_{r_2}^l(t) = U^l_{r_2} \left(
        \mathbf{m}_{r_2}^l(t)
    \right),
\end{empheq}
where $C^l(\cdot)$, $U^l_{r_1}(\cdot)$ and $U^l_{r_2}(\cdot)$ are linear transformation layers for learning common features, and unique features for the \textit{is\_subtopic\_of} and \textit{has\_tag} relation, respectively.

In order to ensure the above separation is successful, we need to keep the distributions of $\mathbf{c}_{r_1}^l(t), \mathbf{c}_{r_2}^l(t)$ as close as possible, and those of $\mathbf{u}_{r_1}^l(t), \mathbf{u}_{r_2}^l(t)$ as far as possible. There are two potential methods. One is to measure the distance between distributions using mutual information. Its drawback is that only the upper and lower bound of mutual information can be estimated. Thus, we adopt the other method, i.e., adversarial training, which is built around a min-max game. A discriminator is used to distinguish two unique features $\mathbf{u}_{r_1}^l(t), \mathbf{u}_{r_2}^l(t)$ while being confused about the common ones $\mathbf{c}_{r_1}^l(t), \mathbf{c}_{r_2}^l(t)$. Formally, we have:
\begin{equation} \label{eq:l_adv}
    \underset{D^l,U^l_{r_1},U^l_{r_2}}{\textmd{min}}\underset{C^l}{\textmd{max}} \ \mathcal{L}^l_{\textmd{adv}} = \mathcal{L}^l_u + \lambda \mathcal{L}^l_c,
\end{equation}
\begin{empheq}[left=\empheqlbrace]{align}
    &\mathcal{L}^l_u = - \textmd{log}\left( D^l \left( \mathbf{u}_{r_1}^l(t) \right) \right) -
        \textmd{log}\left( 1 - D^l \left(\mathbf{u}_{r_2}^l(t) \right) \right), \\
    &\mathcal{L}^l_c = - \textmd{log}\left( D^l \left( \mathbf{c}_{r_1}^l(t) \right) \right) - 
        \textmd{log}\left( 1 - D^l \left(\mathbf{c}_{r_2}^l(t) \right) \right),
\end{empheq}
where $D^l$ is a binary linear classifier served as the discriminator, $\lambda$ is a trade-off parameter, and $\mathcal{L}^l_c, \mathcal{L}^l_u$ are discrimination loss for common and unique features, respectively.

After common-unique separation, we directly aggregate them to derive the output representation $\mathbf{h}^l(t)$ for tag node $t$:
\begin{empheq}[left=\empheqlbrace]{align}
    &\mathbf{c}^l(t) = \textmd{MeanPool} \left( \mathbf{c}_{r_1}^l(t) + \mathbf{c}_{r_2}^l(t) \right), \\
    &\mathbf{u}^l(t) = \textmd{MaxPool} \left( \mathbf{u}_{r_1}^l(t) + \mathbf{u}_{r_2}^l(t) \right), \\
    &\mathbf{h}^l(t) = \frac{1}{2} \left( \mathbf{u}^l(t) + 
        \mathbf{c}^l(t) \right).
\end{empheq}

\subsubsection{Training and Inference} \label{sec:train_inference}
We compute the semantic similarity between a video and all tags as our predictions. Specifically, for each video node $v$ and tag node $t$, the probability whether node $v$ has tag $t$ is estimated as $\hat{s}(v,t)\in\mathbb{R}$:
\begin{equation}
    \hat{s}(v,t) = \textmd{Sigmoid} \left( \mathbf{h}^L(v) \cdot \left( \mathbf{h}^L(t) \right)^T \right).
\end{equation}

During training, we apply the binary cross-entropy (BCE) loss $L_\textmd{tag}$ as tagging loss:
\begin{equation}
    \mathcal{L}_{\textmd{tag}} = \textmd{BCE} \left( y(v,t), \hat{s}(v,t) \right),
\end{equation}
where $y(v,t)$ is the ground truth for whether video $v$ has tag $t$.

We use gradient reversal layer (GRL) \cite{ganin_grl_2015} to implement the min-max game in \autoref{eq:l_adv} for end-to-end training. The final loss $L$ is composed of classification loss and adversarial training loss:
\begin{equation}
    \mathcal{L} = \mathcal{L}_{\textmd{tag}} + \sum_{l=1}^L{\mathcal{L}^l_{\textmd{adv}}}.
\end{equation}

During inference, a new video has no out-coming relations, because it has no tags (no \textit{has\_tag} relations), and only previously-uploaded videos can influence the newly-uploaded ones (no out-coming \textit{is\_followed\_by} relations). Hence the node representation of old nodes can be computed in advance. Thus, when a new video is added to the video-tag network, we only need to calculate its in-coming \textit{is\_followed\_by} messages $\mathbf{m}^l_{r_3}(t)$ to get the video representation. Therefore, our model can achieve inductive inference.

\section{Experiments}

\begin{table}[t]

\caption{Dataset Statistics.}
\vspace{-1em}
\resizebox{\linewidth}{!}{
\begin{tabular}{c|c|c|cc|cc}
\hline
                                   &                                    &                                  & \multicolumn{1}{c|}{\textbf{\#Tags}} & \multicolumn{1}{l|}{\textbf{\#Followers}} & \multicolumn{1}{c|}{\textbf{\#Videos}} & \multicolumn{1}{l}{\textbf{\#Subtopics}} \\ \cline{4-7} 
\multirow{-2}{*}{\textbf{Dataset}} & \multirow{-2}{*}{\textbf{\#Video}} & \multirow{-2}{*}{\textbf{\#Tag}} & \multicolumn{2}{c|}{{\color[HTML]{333333} \textbf{/Video}}}                  & \multicolumn{2}{c}{{\color[HTML]{333333} \textbf{/Tag}}}                      \\ \hline
apparel                            & 245,353                            & 4,817                            & \multicolumn{1}{c|}{5.1}           & 468.4                                   & \multicolumn{1}{c|}{259.8}           & 3.0                                    \\
cosmetics                           & 96,924                             & 3,114                            & \multicolumn{1}{c|}{4.0}           & 101.5                                   & \multicolumn{1}{c|}{124.5}           & 2.2                                    \\
food                               & 108,206                            & 8,640                            & \multicolumn{1}{c|}{4.3}           & 24.6                                    & \multicolumn{1}{c|}{53.8}            & 2.4                                    \\ \hline
\end{tabular}
}

\label{table:dataset_stat}
\end{table}

\subsection{Dataset}
\subsubsection{Data Collection and Preparation}
We obtained a large-scale and real-world micro-video dataset from three verticals (apparel, cosmetics, and food) on Kuaishou platforms. We randomly selected 500,000 micro-videos on the featured page from July 14 to August 14, 2021. Their tags are obtained from a variety of human-based signals (e.g., query-watch) and the online video tagging system which leverages only video textual descriptions. The follow relations among video creators is collected on August 14, and we filtered robot users along with their videos. We excluded too short (< 1 sec) or too long (> 60 secs) micro-videos, removed the non-visual tags, and filtered videos with less than 2 tags.

To quantify the dataset quality, we manually evaluated 1000 randomly sampled micro-videos, and the tagging accuracy and recall is 91.3\% and 66.2\%, respectively. Our dataset is split into 3 partitions: train (80\%), validate (10\%), and test (10\%).

\begin{table*}[t]

\caption{Performance comparison with the baselines.}
\vspace{-1em}
\resizebox{\textwidth}{!}{
\begin{tabular}{c|ccccc|ccccc|ccccc}
\hline
\multirow{2}{*}{Model} & \multicolumn{5}{c|}{Apparel Dataset}                                               & \multicolumn{5}{c|}{Cosmetics Dataset}                                              & \multicolumn{5}{c}{Food Dataset}                                                   \\ \cline{2-16} 
                       & mAP            & P@1            & P@3            & R@5            & R@10           & mAP            & P@1            & P@3            & R@5            & R@10           & mAP            & P@1            & P@3            & R@5            & R@10           \\ \hline
Vanilla                & 59.89          & 89.22          & 74.97          & 57.12          & 64.68          & 46.33          & 74.30          & 45.49          & 47.23          & 56.06          & 45.98          & 60.13          & 47.78          & 50.57          & 59.60          \\
ML-GCN                 & 61.49          & 91.24          & 76.75          & 57.97          & 65.53          & 48.03          & 77.20          & 46.42          & 48.15          & 57.02          & 47.54          & 63.94          & 49.65          & 51.75          & 59.97          \\
NeXtVLAD               & 63.14          & 91.90          & 77.44          & 58.86          & 66.98          & 49.78          & 77.93          & 47.93          & 49.58          & 58.92          & 48.49          & 64.92          & 50.51          & 52.43          & 60.93          \\
CMA                    & 63.04          & 91.51          & 77.29          & 58.85          & 66.99          & 50.24          & 78.60          & 48.29          & 50.04          & 59.41          & 49.28          & 65.54          & 50.91          & 53.09          & 61.65          \\
MALL-CNN               & 63.31          & 91.67          & 77.95          & 59.07          & 67.20          & 50.46          & 78.93          & 48.25          & 50.15          & 59.83          & 49.58          & 66.00          & 50.89          & 53.32          & 61.99          \\ \hline
TagGNN                 & 63.44          & 93.15          & 78.23          & 59.13          & 66.82          & 52.39          & 84.80          & 50.32          & 51.64          & 60.37          & 50.92          & 74.49          & 53.58          & 53.09          & 60.05          \\
HAN                    & 61.83          & 92.17          & 77.06          & 58.00          & 65.76          & 51.72          & 84.87          & 49.54          & 51.31          & 59.99          & 50.76          & 75.32          & 53.53          & 53.02          & 59.79          \\
HGT                    & 63.85          & 92.35          & 78.25          & 59.43          & 67.58          & 54.50          & 86.62          & 51.78          & 53.61          & 62.62          & 51.56          & 76.22          & 54.38          & 53.53          & 60.04          \\
SimpleHGN              & 63.16          & 91.80          & 77.99          & 59.09          & 67.10          & 53.68          & 86.28          & 51.17          & 52.80          & 61.62          & 52.76          & 77.27          & 55.21          & 54.19          & 61.33          \\ \hline
\textbf{Ours}          & \textbf{65.05} & \textbf{93.74} & \textbf{79.07} & \textbf{59.89} & \textbf{68.22} & \textbf{56.55} & \textbf{88.16} & \textbf{53.18} & \textbf{54.98} & \textbf{64.55} & \textbf{56.93} & \textbf{82.16} & \textbf{58.94} & \textbf{57.11} & \textbf{63.99} \\ \hline
\end{tabular}
}

\label{table:comp_results}

\end{table*}

\subsubsection{Tag Ontology Construction}

\begin{figure}[t]
    \centering
    \includegraphics[width=0.6\linewidth]{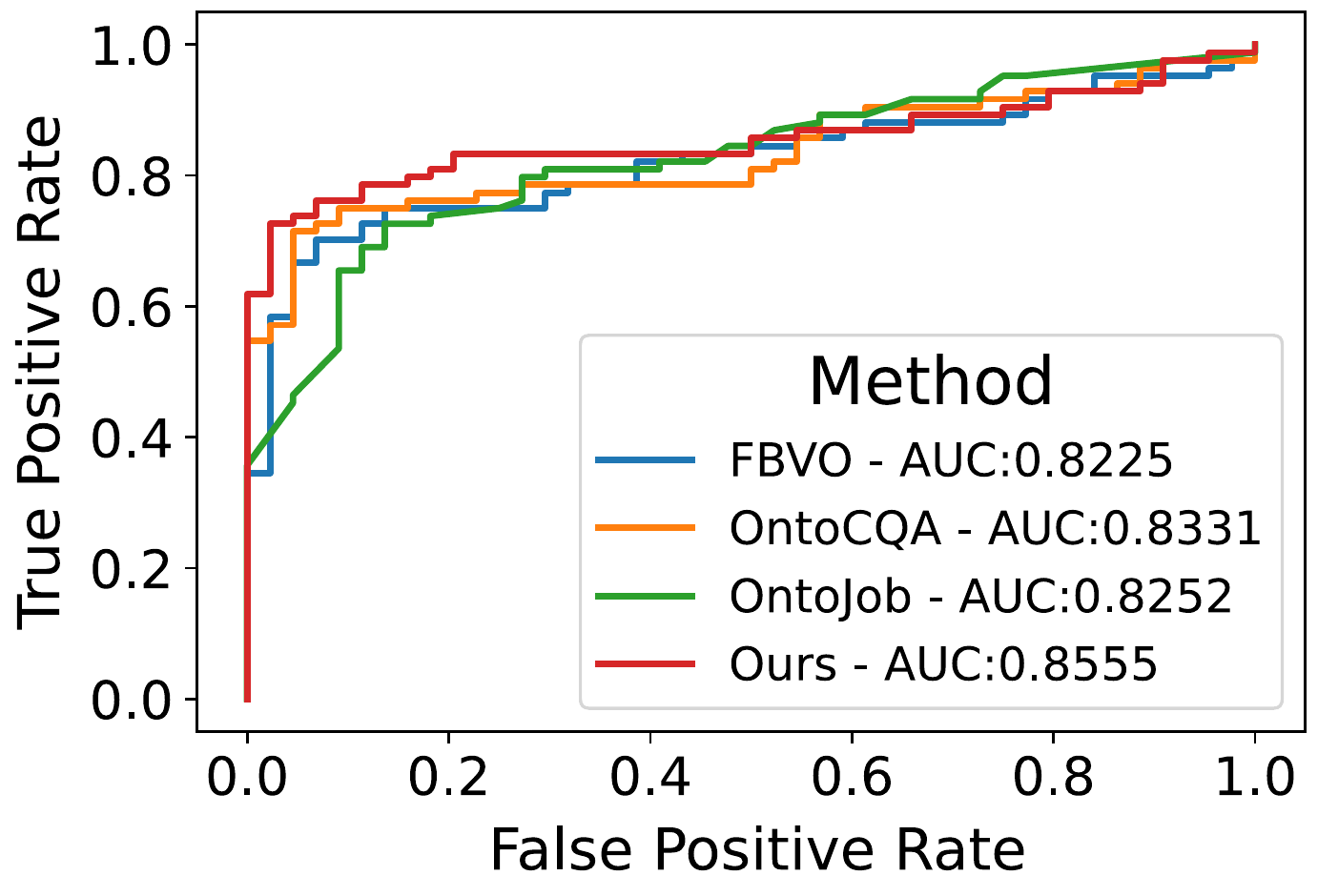}
    \vspace{-1em}
    \caption{ROC of subtopic discovery}
    \label{fig:onto_cons_roc}
    \vspace{-2em}
\end{figure}

We constructed the tag ontology according to our proposed method in \autoref{sec:ontology_construction}. We annotated 1,000 out of 768,798 co-occurred tag pairs. The two thresholds $\delta_r,\epsilon_r$ are defined according to the precision and recall of subtopic discovery. Concretely, we chose $\delta_r$ where the precision is 90\%, and $\epsilon_r$ where the recall is 90\%. As illustrated in \autoref{fig:onto_cons_roc}, we measured the construction performance of subtopic discovery, and our method outperforms the rule-based method FBVO \cite{fang_folksonomy-based_2016}, unsupervised method OntoJob \cite{vrolijk_ontojob_2022} and OntoCQA \cite{Suryamukhi_modeling_2021} in a clear margin. Our constructed ontology covers 96.67\% of tags, and only 2.94\% of tags are covered through $\epsilon_r$ relaxation.

\subsubsection{Dataset Statistics}

We presented the statistics of our dataset in \autoref{table:dataset_stat}. Their diverged statistics lead to different results in the following experiments. The apparel dataset has the richest visual information (the \#videos per tag), and the food dataset has the lowest social network and tag ontology density (the least \#followers per video and the second least \#subtopics per tag).

\subsection{Experimental Settings}

\subsubsection{Experimental Details}

The tag word embedding is derived from the last hidden layer of BERT. The video frame embedding is acquired from TSN backboned by SwinTransformer. 
In every GNN layer, we sample 4 neighbors for each type of relations following the practice of Deep Graph Library \cite{wang2019dgl} because of the GPU memory limit. The number of layers $L$ is set to 2. We set the batch size to be 1024, and used AdamW optimizer with a learning rate of 0.0005. We set feature and edge dropout to 0.2. Following \citeauthor{ganin_grl_2015} \cite{ganin_grl_2015} to stabilize the adversarial training, we gradually changed $\lambda$ from 0 to $\lambda_0$ according to the schedule:
$\lambda = \lambda_0 \left( 2/(1+exp(-\gamma p))-1 \right)$,
where $p$ linearly changes from 0 to 1 in the training process. The $\gamma$ and $\lambda_0$ was set to 20 and 0.0005 for all datasets, respectively. All the settings are applied equally to all baseline methods.

\subsubsection{Evaluation Metrics} We adopted three common metrics for multi-label classification: mean Average Precision (mAP), Precision@K (P@K), and Recall@K (R@K). The P@K and R@K measure the tagging performance of top K tags, while mAP measures the overall ranking results of all tags.

\subsection{Baselines}
In order to verify our RADAR model, we compared it with the following baselines:
\begin{itemize}
    \item Video tagging methods: ML-GCN \cite{ml_gcn}, NeXtVLAD \cite{nextvlad, wu_tencent_mmads_1st_2021}, CMA \cite{cma}, and MALL-CNN \cite{li_mall_2022}. 
    We compared them to demonstrate the effectiveness of incorporating extra social networks and tag graph information. Note that NeXtVLAD, CMA, and MALL-CNN have sequence-level feature $\mathbf{h}\in\mathbb{R}^{F\times d}$ as input, with an extra aggregation module instead of meaning pooling in our method.
    \item Heterogeneous GNN methods: HAN \cite{han}, TagGNN \cite{taggnn}, HGT \cite{hgt}, and Simple-HGN \cite{simple_hgn} are four representative methods on heterogeneous GNN, and Simple-HGN is the state-of-the-art. 
    We compared our RADAR with them to demonstrate the effectiveness of our RADAR model.
\end{itemize}

\subsection{Performance Comparison}

Compared with baselines, our results are summarized in \autoref{table:comp_results}. By analyzing this table, we gained the following observations:
\begin{itemize}
    \item The heterogeneous GNN methods generally outperform the video tagging baselines. Their main difference is that tagging baselines better leverage video frame features, and GNN methods leverage tag ontology and extra social network. The improvement indicates that for micro-videos, incorporating social influence and tag relations is more effective than modeling frame features.
    \item Our proposed model RADAR outperforms all heterogeneous GNN methods consistently. Compared with the best-performing baselines, RADAR obtains relative mAP gains with 1.9\% in Apparel, 3.8\% in Cosmetics, and 7.9\% in Food dataset. The improvement demonstrates the superiority of RADAR in video tagging.
    \item Simple-HGN is the only heterogeneous GNN baseline whose relevant rank differs among datasets. Notably, its performance is the best in Food while the second-worst in Apparel dataset within all heterogeneous GNN baselines. Coincidentally, this relevant rank is opposite to the datasets' social network and tag ontology density. The denser \textit{is\_followed\_by} and \textit{is\_subtopic\_of} relations are, the worse Simple-HGN performs. Considering that Simple-HGN is the start-of-the-art method for general heterogeneous GNN, we attributed this phenomenon to the fact that video tagging is quite different from the previous tasks. In video tagging, The video feature from frames and tag feature from language models are two modalities with large semantic gaps. This phenomenon also demonstrates the necessity of our adversarial aggregation network which aggregates the visual and linguistic information.
\end{itemize}

\subsection{Ablation Studies}

In this section, we carried out several experiments to further analyze the effectiveness of our model, as reported in \autoref{table:ablation}. Concretely, we first explored the contributions of different relations in the video-tag network. We then analyzed the effectiveness of each component including Gated Graph Transformer (GGT) and Adversarial Aggregation Network (AAN).

\subsubsection{Ablation of different relations} We compared our model with the following variants: 1) \textbf{w/o followed by}, removing \textit{is\_followed\_by} relations; 2) \textbf{w/o subtopic}, removing \textit{is\_subtopic\_of} relations; and 3) \textbf{w/o subtopic \& followed by}, removing both relations. 

\begin{table*}[t]

\caption{Ablation study on the effectiveness of the two relations, gated graph transformer, and adversarial aggregation network.}
\vspace{-1em}
\resizebox{\textwidth}{!}{
\begin{tabular}{c|ccccc|ccccc|ccccc}
\hline
\multirow{2}{*}{Model}                           & \multicolumn{5}{c|}{Apparel Dataset}  & \multicolumn{5}{c|}{Cosmetics Dataset} & \multicolumn{5}{c}{Food Dataset}      \\ \cline{2-16} 
                                                 & mAP   & P@1   & P@3   & R@5   & R@10  & mAP   & P@1   & P@3   & R@5   & R@10  & mAP   & P@1   & P@3   & R@5   & R@10  \\ \hline
w/o followedby                                   & 62.28 & 91.49 & 76.97 & 58.49 & 66.34 & 48.68 & 77.71 & 47.11 & 48.72 & 58.00 & 47.37 & 63.89 & 49.87 & 51.78 & 60.26 \\
w/o subtopic                                     & 63.60 & 93.56 & 78.43 & 59.17 & 66.96 & 53.17 & 86.23 & 50.93 & 52.43 & 60.98 & 53.27 & 79.39 & 56.37 & 54.67 & 60.74 \\
w/o subtopic \& followedby                       & 62.17 & 91.45 & 76.97 & 58.39 & 66.18 & 48.33 & 77.35 & 46.88 & 48.59 & 57.57 & 46.48 & 63.33 & 48.82 & 51.38 & 59.35 \\ \hline
w/o gated residual                               & 64.20 & 93.32 & 78.54 & 59.52 & 67.66 & 55.20 & 86.77 & 52.06 & 54.06 & 63.27 & 55.65 & 80.90 & 57.74 & 56.18 & 62.94 \\
w/ mutual attention                              & 64.12 & 93.34 & 78.55 & 59.44 & 67.61 & 55.32 & 87.35 & 52.14 & 54.10 & 63.35 & 55.69 & 81.00 & 57.66 & 56.34 & 63.24 \\ \hline
w/ concatenation                                 & 63.82 & 93.34 & 78.50 & 59.24 & 67.16 & 53.48 & 86.57 & 51.20 & 52.58 & 61.32 & 52.64 & 78.25 & 55.67 & 54.20 & 60.50 \\
w/ attention                                     & 64.76 & 93.48 & 78.97 & 59.78 & 67.89 & 55.27 & 88.01 & 52.30 & 53.89 & 62.86 & 55.47 & 81.64 & 58.13 & 56.11 & 62.45 \\
w/ LMF                                           & 64.36 & 93.25 & 78.98 & 59.63 & 67.72 & 55.00 & 87.51 & 52.25 & 53.67 & 62.42 & 54.85 & 81.24 & 57.64 & 55.69 & 61.71 \\
w/o $\boldsymbol{L}_{adv}$                       & 63.87 & 93.21 & 78.60 & 59.38 & 67.28 & 54.31 & 86.57 & 51.83 & 53.47 & 62.26 & 56.57 & 82.08 & 58.78 & 56.88 & 63.44 \\ \hline
w/ bidirectional edges                           & 65.26 & 93.94 & 79.13 & 60.09 & 68.35 & 56.86 & 88.06 & 53.34 & 55.12 & 64.65 & 57.19 & 82.27 & 58.98 & 57.35 & 64.32 \\
Ours                                             & 65.05 & 93.74 & 79.07 & 59.89 & 68.22 & 56.55 & 88.16 & 53.18 & 54.98 & 64.55 & 56.93 & 82.16 & 58.94 & 57.11 & 63.99 \\ \hline
\end{tabular}
}

\label{table:ablation}
\end{table*}

The result of \textbf{w/o followed by} is consistently lower than of \textbf{w/o subtopic} in three datasets, suggesting that the social network information contributes more than tag ontology. 

Note that if we remove \textit{is\_followed\_by} from our full model, the performance drop is more significant than we remove \textit{is\_followed\_by} from \textbf{w/o subtopic}. 
In another view, the addition of \textit{is\_followed\_by} relations will bring larger improvement when the \textit{is\_subtopic\_of} relations exist. Therefore, the combination of the two relations is not the naive summation of the two relations. This is evidence that the two relations have complementary information, and thus the separation of common and unique information in the AAN module is necessary.

\subsubsection{Ablation of GGT} We replaced the gated residual network in GGT with a residual network to be the variant \textbf{w/o gated residual}, and replaced the separate attention with mutual attention to be the variant \textbf{w/ mutual attention}. Compared with our full model, the performance of \textbf{w/o gated residual} and \textbf{w/ mutual attention} consistently dropped in all datasets. This indicates that it is important to filter irrelevant information for modeling Behavior Spread, and separate attention is a useful design under the modality gap.
 
\subsubsection{Ablation of AAN} To validate the impact of our proposed AAN, we conducted a series of experiments by introducing four variants: 1) \textbf{w/ concatenation}, replacing AAN with vector concatenation and a linear transformation layer, which is the simplest multi-modal aggregation method; 2) \textbf{w/ attention}, replacing AAN with attention module in \cite{han}; 3) \textbf{w/ LMF}, replacing AAN with Low-rank Multi-modal Fusion (LMF) \cite{liu_lmf_2018}, which is the best multi-modal vector aggregation method; and 4) \textbf{w/o $\boldsymbol{L}_{adv}$}, removing the adversarial loss $L_{adv}$. 

From \autoref{table:ablation}, we observed that our method outperforms all the multi-modal vector aggregation methods consistently. The improvement is not owing to the increased parameters, because \textbf{w/ LMF} has the most parameters. Compared with \textbf{w/o $\boldsymbol{L}_{adv}$}, we attributed the improvement to the separation of common and unique information in the visual and linguistic messages.

\subsubsection{Bidirectional edges}
We also explored the effect of bidirectional edges. Specifically, in the variant \textbf{w/ bidirectional edges}, we added \textit{is\_supertopic\_of} and \textit{has\_video} relations to the network. We also applied AAN to video nodes since they have two types of inbound messages. Note that we did not add \textit{follows} relations, because we hope the new videos do not send messages to the original graph for inductive learning. Although this variant slightly improves the performance, we do not adopt this setting considering the extra computational burden.

\subsection{Sensitivity Analysis}

We investigated how different choices of hyper-parameters affect performance. We explored two main parameters: the number of GNN layer $L$ and the maximal weight of unique loss $\lambda_0$.

As illustrated in \autoref{fig:sensitivity_analysis}, our model performs best when it has two GNN layers, which is common in graph neural networks due to over-smoothing. Our model achieves the best performance when $\lambda_0$ is around 0.0005 to 0.001, which indicates an appropriate balance between unique and common loss. Note that $\lambda_0=0$ is different from \textbf{w/o $\boldsymbol{L}_{adv}$}. Although both of them have no common information, the discriminator $D^l$ in the latter variant is removed, so it has no unique information as well. We observed that \textbf{w/o $\boldsymbol{L}_{adv}$} will cause a larger performance drop compared with setting $\lambda_0=0$. This indicates that the unique information is more important.

\begin{figure}[t]
    \centering
    
    \begin{subfigure}{.33\linewidth}
        \centering
        \includegraphics[width=\linewidth]{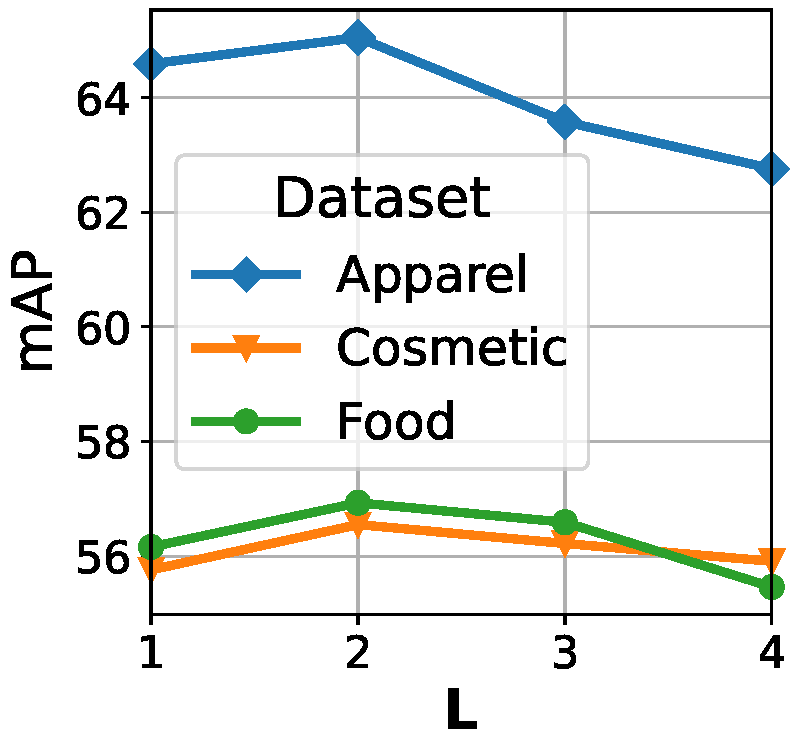}  
        \label{fig:sensi_L}
    \end{subfigure}
    \begin{subfigure}{.655\linewidth}
        \centering
        \includegraphics[width=\linewidth]{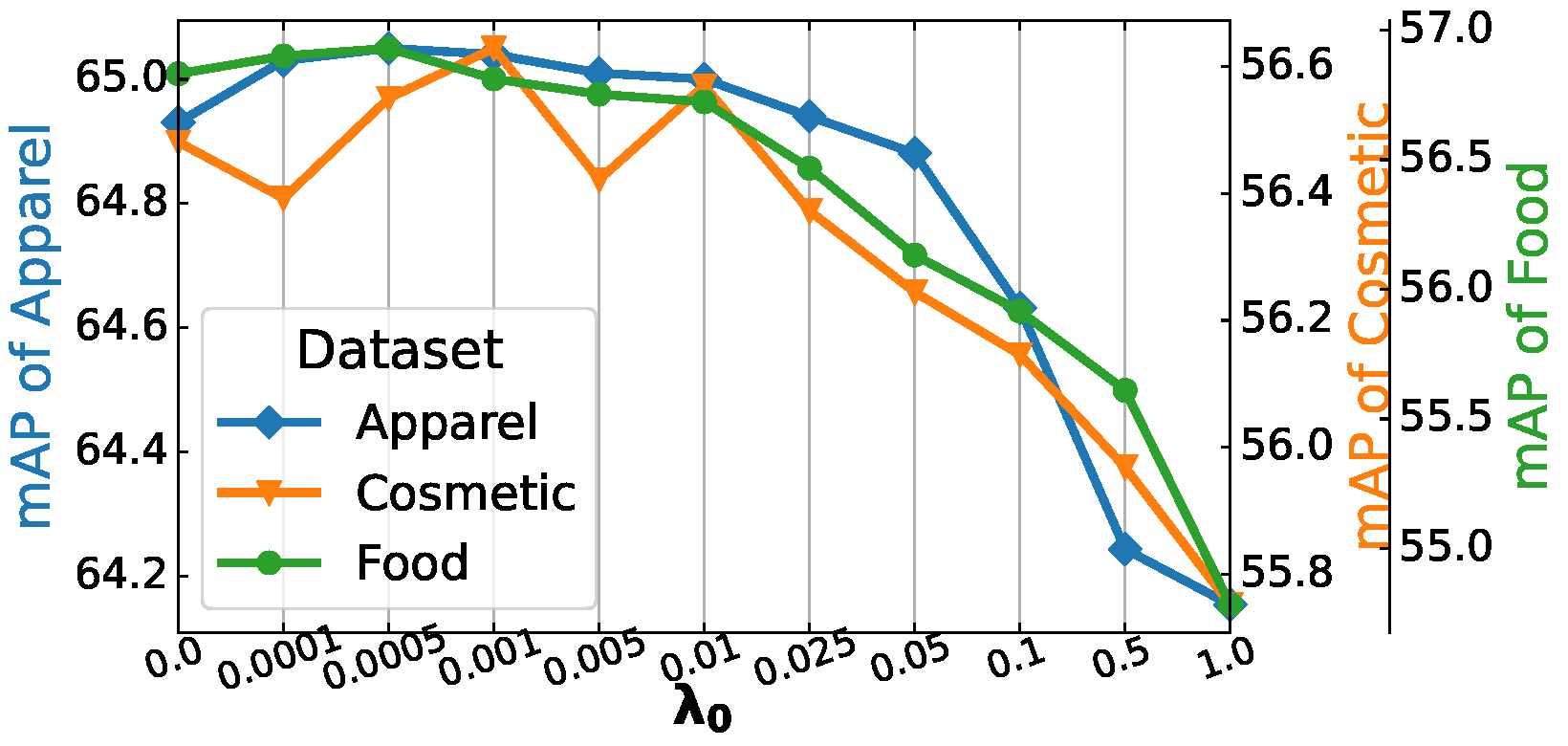}  
        \label{fig:sensi_lambda0}
    \end{subfigure}
    \vspace{-2em}
    \caption{Sensitivity analysis}
    \label{fig:sensitivity_analysis}
\end{figure}

\section{Conclusion and Future Work}
In this work, we jointly incorporate social influence and tag relations into a user-video-tag network to enhance micro-video tagging.
To construct tag relations, we build a tag ontology based upon tag statistics and graphical constraints in a semi-supervised manner. 
To derive a better node representation in video-tag network, we present RADAR, a heterogeneous GNN composed of a gated graph transformer and adversarial aggregation network. Extensive experimental results on three datasets have demonstrated the superiority of RADAR, and the effectiveness of its two components.

In the future, 
we are going to generalize our method to handle more complicated tag knowledge structures. 
As to the Behavior Spread modeling, we plan to model the temporal dynamics of users’ social influence.

\section{Acknowledgements}
This work is supported by the National Natural Science Foundation of China, No.: 62176137, No.:U1936203, and No.: 62006140; the Shandong Provincial Natural Science and Foundation, No.: ZR2020QF106; Beijing Academy of Artificial Intelligence(BAAI); Kuaishou Technology.

\endgroup

\clearpage
\bibliographystyle{ACM-Reference-Format}
\bibliography{citations}

\end{document}